\documentclass[]{spie}  
\usepackage[nonumberlist,style=long,acronym]{glossaries}

\newacronym{braed}{BrAED}{Broadband Asynchronous Entanglement Distribution}
\newacronym{qnet}{QNET}{Quantum Network with Evolving Topology}
\newacronym{qn}{QN}{Quantum Network}
\newacronym{qi}{QI}{Quantum Internet}
\newacronym{es}{ES}{Entanglement Source}
\newacronym{pd}{PD}{Polarisation Definition}
\newacronym{lc}{LC}{Logical Channel}

\newacronym{dwdm}{DWDM}{Dense Wavelength Division Multiplexing}
\newacronym{itu}{ITU}{International Telecommunication Union}
\newacronym{qroadm}{q-ROADM}{quantum-enabled Reconfigurable Optical Add-Drop Multiplexer}
\newacronym{pps}{PPS}{Photon Pair Source}
\newacronym{rq}{Rq}{Quantum Receiver}
\newacronym{tx}{Tx}{Classical Communication Transmitter}
\newacronym{rx}{Rx}{Classical Communication Receiver}
\newacronym{fpc}{FPC}{Fibre Polarisation Controller}
\newacronym{mpc}{MPC}{Motorised Fibre Polarisation Controller}
\newacronym{os}{OS}{Optical Switch}
\newacronym{fbs}{FBS}{Fibre Beam Splitter}
\newacronym{mux}{Mux}{Multiplexer}
\newacronym{demux}{Demux}{Demultiplexer}
\newacronym{hcf}{HCF}{Hollow Core Fibre}
\newacronym{smf}{SMF}{Single Mode Fibre}
\newacronym{fc}{FC}{Fibre Collimater}

\newacronym{spdc}{SPDC}{Spontaneous Parametric Down Conversion}
\newacronym{ppln}{PPLN}{Periodically Poled Lithium Niobat}
\newacronym{mg-ppln}{Mg-PPLN}{Magnesium-doped Periodically Poled Lithium Niobat}
\newacronym{qpm}{QPM}{quasi-phasematching}

\newacronym{voa}{VOA}{Variable Optical Attenuator}

\newacronym{shg}{SHG}{Second Harmonic Generation}
\newacronym{ed}{ED}{Entanglement Distribution}
\newacronym{he}{HE}{Heralding Efficiency}
\newacronym{spd}{SPD}{Single-Photon Detector}
\newacronym{snspd}{SNSPD}{Superconducting Nanowire Single-Photon Detector}

\newacronym{cw}{CW}{Coincidence Window}
\newacronym{fwhm}{FWHM}{Full Width at Half Maximum}

\newacronym{qkd}{QKD}{Quantum Key Distribution}
\newacronym{sk}{SK}{Secret Key}
\newacronym{skr}{SKR}{Secret Key Rate}
\newacronym{aeskr}{AE-SKR}{Average Effective Secret Key Rate}
\newacronym{qber}{QBER}{Quantum Bit Error Rate}
\newacronym{qc}{QC}{Quantum Communication}

\newacronym{hv}{HV}{Horizontal-Vertical}
\newacronym{da}{DA}{Diagonal-Anti-Diagonal}
\newacronym{h}{H}{Horizontal}
\newacronym{v}{V}{Vertical}
\newacronym{d}{D}{Diagonal}
\newacronym{a}{A}{Anti-Diagonal}

\newacronym{bbm92}{BBM92}{Bennett-Brassard-Mermin 1992 QKD protocol}

\newglossaryentry{polarisation}{
        name=Polarisation,
        description={Polarisation is the property of light that defines which plane that transverse waves are oscillating in. In this work we are using Linearly polarised light and this definition is limited to this basis}
}
\newglossaryentry{Quantum Technology}{
        name=Quantum Technology,
        description={Technology that leverages the fundamental stated of quantum system to preform tasks that would otherwise be impossible or inefficient on classical systems. Examples include Quantum Computation, Quantum Sensing, and Quantum Key Distribution}
}
\newglossaryentry{Quantum Network}{
        name=Quantum Network,
        description={A network that shares quantum states between nodes, rather than sending bright classical signals}
}
\newglossaryentry{Quantum Computing}{
        name=Quantum Computing,
        description={A type of computation that harnesses the properties of quantum states, such as entanglements and superposition, to preform tasks}
}
\newglossaryentry{Quantum Sensing}{
        name=Quantum Sensing,
        description={A type of Sensing.......}
}
\newglossaryentry{Co-Propagation}{
        name=Co-Propagation,
        description={The propagation of two distinct signals of light, down the same fibre, in the same direction}
}
\newglossaryentry{Counter-Propagation}{
        name=Counter-Propagation,
        description={The propagation of two distinct signals of light, down the same fibre, in opposite directions}
}
\newglossaryentry{coex}{
        name=Co-Existence,
        description={The propagation of both classical power light and quantum level light in the same fibre simultaneously}
}
\newglossaryentry{Scattering}{
        name=Scattering,
        description={The scattering of photons either by Raman, in-elastic, scattering, or Rayleigh, elastic, scattering of photons in the optical fibre. This includes both Stokes and anti-Stokes scattering}
}
\newglossaryentry{Dispersion}{
        name=Dispersion,
        description={The spreading out of the probable arrival time of photons as they travel along an optical fibre. This includes effects from the waveguide dispersion, where light travels both in the core and the cladding of the fibre, and the material dispersion, due to the difference in refractive index for photons of different frequencies}
}
\newglossaryentry{C-Band}{
        name=C-Band,
        description={In Optical communication, this refers to the wavelength range from $1530$\,nm to $1565$\,nm}
}
\newglossaryentry{O-Band}{
        name=O-Band,
        description={In Optical communication, this refers to the wavelength range from $1260$\,nm to $1360$\,nm}
}
\newglossaryentry{Entanglement Swapping}{
        name=Entanglement Swapping,
        description={Teleproting Quantum states between pairs of entangled photons, such that in the end one photon from each pair survive and are entangled with each other}
}

\newglossaryentry{test}{
        name=Test,
        description={... Test ... Test ... Test ... Test ... Test ... Test ... Test ... Test ... Test ... Test ... Test ... Test ... Test ... Test ... Test ... Test ... Test ... Test ... Test ... Test ... Test ... Test ... Test ... Test ... Test ... Test ... Test ... Test ... Test ... Test ... Test ... Test ... Test ... Test ... Test ... Test ... Test ... Test ... Test ... Test ... Test ... Test ... Test ... Test ... Test ... Test ... Test ... Test ... Test ... Test ... Test ... Test ... Test ... Test ... Test ... Test ... Test ... Test ... Test ... Test ... Test ... Test ... Test ... Test ... Test ... Test ... Test ... Test ... Test ... Test ... Test ... Test ... Test ... Test ... Test ... Test ... Test ... Test ... Test ... Test ... Test ... Test ... Test ... Test ... Test ... Test ... Test ... Test ... Test ... Test ... Test ... Test ... Test ... Test ... Test ... Test ... Test ... Test ... Test ... Test ... Test ... Test ... Test ... Test ... Test ... Test ... Test ... Test ... Test ... Test ... Test ... Test ... Test ... Test ... Test ... Test ... Test }
}
\usepackage{glossary-mcols}
\usepackage[T1]{fontenc} 

\usepackage{amsmath,amsfonts,amssymb}
\usepackage{graphicx}
\usepackage[colorlinks=true, allcolors=blue]{hyperref}

\title{Entanglement distribution quantum networking within deployed telecommunications fibre-optic infrastructure}

\author[a, b]{\small M.~J.~Clark}  
\author[c]{O.~Alia}                 
\author[c]{R.~Wang}                 
\author[c]{S.~Bahrani}              
\author[e]{M.~Perani\'{c}}          
\author[d]{D.~Aktas}                
\author[c]{G.~T.~Kanellos}          
\author[e]{M.~Loncaric}             
\author[e]{\v{Z}.~Samec}            
\author[e]{A.~Radman}               
\author[e]{M.~Stipcevic}            
\author[c]{R.~Nejabati}             
\author[c]{D.~Simeonidou}           
\author[a]{J.~G.~Rarity}            
\author[a]{S.~K.~Joshi}             
\affil[a]{\small Quantum Engineering Technology Labs, NSQI, University of Bristol, 5 Tyndall Ave., Bristol, England, BS8 1FD}
\affil[b]{\small Quantum Engineering Centre of Doctoral training, NSQI, University of Bristol, 5 Tyndall Ave., Bristol, England, BS8 1FD}
\affil[c]{High Performance Networking, University of Bristol, 75 Woodland Rd, Bristol, England, BS8 1UB}
\affil[d]{Research Center for Quantum Information, Institute of Physics, Slovak Academy of Sciences, Dúbravská Cesta 9, 84511 Bratislava, Slovaki}
\affil[e]{Center of excellence for Advanced Materials and Sensing Devices, Ru\dj er Bo\v{s}kovi\'{c} Institute, Bijenička cesta 54, HR-10000 Zagreb, Croatia}

\authorinfo{Further author information: (Send correspondence to M. J. Clark)\\M. J. Clark: E-mail: mj.clark@bristol.ac.uk
}

\begin{document} 
\maketitle

\begin{abstract}
Quantum networks have been shown to connect users with full-mesh topologies without trusted nodes\cite{Joshi2020ANetwork, Wang2022ANetwork}. 
We present advancements on our scalable polarisation entanglement-based quantum network testbed, which has the ability to perform protocols beyond simple quantum key distribution. \cite{Huang2022ExperimentalNetwork,Pelet2022UnconditionallyNetwork}
Our approach utilises wavelength multiplexing, which is ideal for quantum networks across local metropolitan areas due to the ease of connecting additional users to the network without increasing the resource requirements per user.
We show a 10 user fully connected quantum network with metropolitan scale deployed fibre links, demonstrating polarisation stability and the ability to generate secret keys over a period of $10.8$\,days with a network wide average-effective secret key rate of $3.38$\,bps.
\end{abstract}


\glsresetall

\section{INTRODUCTION}
\label{sec:intro}  
In the future landscape of \gls{Quantum Technology} a method of interconnecting different quantum systems will be required.
The prevailing direction is towards a \gls{qi} \cite{Wehner2018QuantumAhead}, constructed by interconnecting \glspl{qn}.
\glspl{qn} have been demonstrated in both metropolitan scale \cite{Chung2022DesignNetwork,Chen2010MetropolitanNetwork} and long distance configurations \cite{Tang2022Free-runningDistribution,Chen2021AnKilometres}, however most have focused solely on \gls{qkd} based networks.
This has been the case as the most near term application of a \gls{qi} is \gls{qkd}. 
Methods for building \glspl{qn} beyond simple \gls{qkd} have been explored and in general are based on \gls{ed}.
Such \glspl{qn} have mainly used photon pair sources to distribute entanglement on a metropolitan scale \cite{Joshi2020ANetwork} and across long distance \cite{Neumann2022ContinuousLink}.
Some have used single photons from local quantum systems to entangle multiple systems over distance \cite{Pompili2021RealizationQubits}.
Progress has been made in generating multi-partite entangled photon states \cite{Avis2022AnalysisNode}, but there are still challenges in producing a scalable network while connect many users together.

As \gls{Quantum Technology} advances a scalable method of distributing entanglement, that allows for multiple simultaneous use cases, is required.
Developing \gls{ed} \glspl{qn} for a metropolitan scale area with minimised resource overheads is essential, while still allowing for a system to be scaled up to a larger set of users.
The \gls{qn} discussed below is a based on \gls{ed} and allows for protocols beyond simple \gls{qkd}. \cite{Huang2022ExperimentalNetwork,Pelet2022UnconditionallyNetwork}

\section{THE QUANTUM NETWORK}
\label{sec:source}
\begin{figure}[t]
    \centering
    \includegraphics[width = \textwidth]{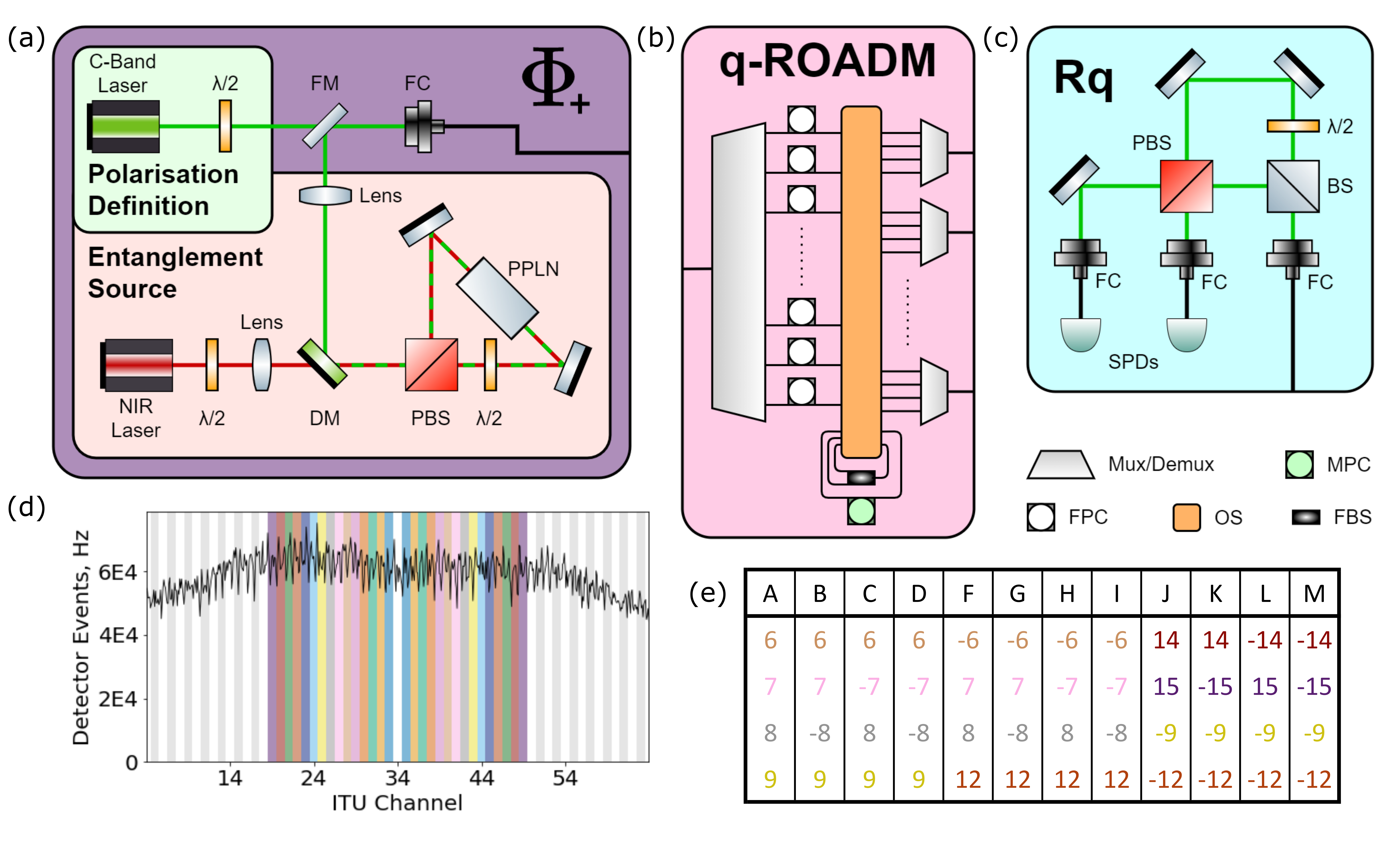}
    \caption{
    \textbf{(a-c)} show the main components of the \acrlong{qn}. 
    \textbf{(a)} is comprised of the \acrlong{es} which produces the entangled photon pairs and the \acrlong{pd} which allows for the definition of polarisation bases to be shared amoungst users. 
    \textbf{(b)} is the \acrlong{qroadm} that separates the source signal into $100$\,GHz ITU channels, routes them, and combines them into a single fibre per user. This systems includes 1-to-4 beam splitters on all demultiplexed channels where $\left| LC \right| \geq 6$. Here FPC is a fibre polarisation controller, MPC is a motorised fibre polarisation controller, OS is an optical switch, FBS is a fibre beams splitter, and a (De)Mux is a wavelength (de)multiplexer. 
    \textbf{(c)} is the \acrlong{rq} that allow users to measure the single photon signal. Here $\lambda/2$ is a half-wave plate, DM is a dichroic mirror, BS is a beam-splitter, PBS is a polarising beam-splitter, PPLN is a magnesium doped periodically poled lithium niobate crystal, FM is a motorised flip mirror, FC is a fibre collimater, and SPD is a single-photon detector. 
    \textbf{(d)} is the spectrum of the \acrlong{es}, where the used \acrlong{itu} channels are shown by bars. Matching channel colour shows entanglement between channels. 
    \textbf{(e)} shows the assignments of \acrlong{lc} to users, where channel colour matches \textbf{(d)}.
    }
    \label{fig:Source}
\end{figure}

The \gls{es} of the \gls{qn}, shown in Figure \ref{fig:Source}\,(a), is a Sangac Interferometer that produces entangled photon pairs through type-0 \acrlong{spdc}. 
A continuous wave $775.06$\,nm laser is used to produce spectrally non-degenerate 
photon pairs with a central wavelength of $1550.12$\,nm. 
Figure \ref{fig:Source}\,(d) shows the frequency spectrum of the \gls{es} output, and how that light is cut into \gls{itu} channels by the \gls{qroadm}, as shown in Figure \ref{fig:Source}\,(b). 
The central frequency of the output is at $193.4$\,THz, corresponding to \gls{itu} channel 34, such that the entanglement is shared pairwise around this channel. 
We label \gls{itu} channels 19 to 49 as \gls{lc} $-15$ to $+15$ respectively.

\glspl{lc} then share entanglement with their negative counterpart. 
It is key to note here that all \gls{itu} channels, in the range \gls{itu} $19-28$ and \gls{itu} $40-49$, have 1-to-4 beam splitters applied to them inside the \gls{qroadm}, such that receiving one \gls{lc} allows a user to communicate with up to 4 other users. 
This allows a users to receive one \gls{lc} in this configuration, where 4 would be required in a configuration without splitting. 
Although this increases the loss on the links by $6$\,dB, it reduces the resource requirement of the \gls{es} and the \gls{qroadm} by requiring fewer \gls{itu} channels to connect all users.

\begin{figure}[ht]
    \centering
    \includegraphics[width = 0.7\textwidth]{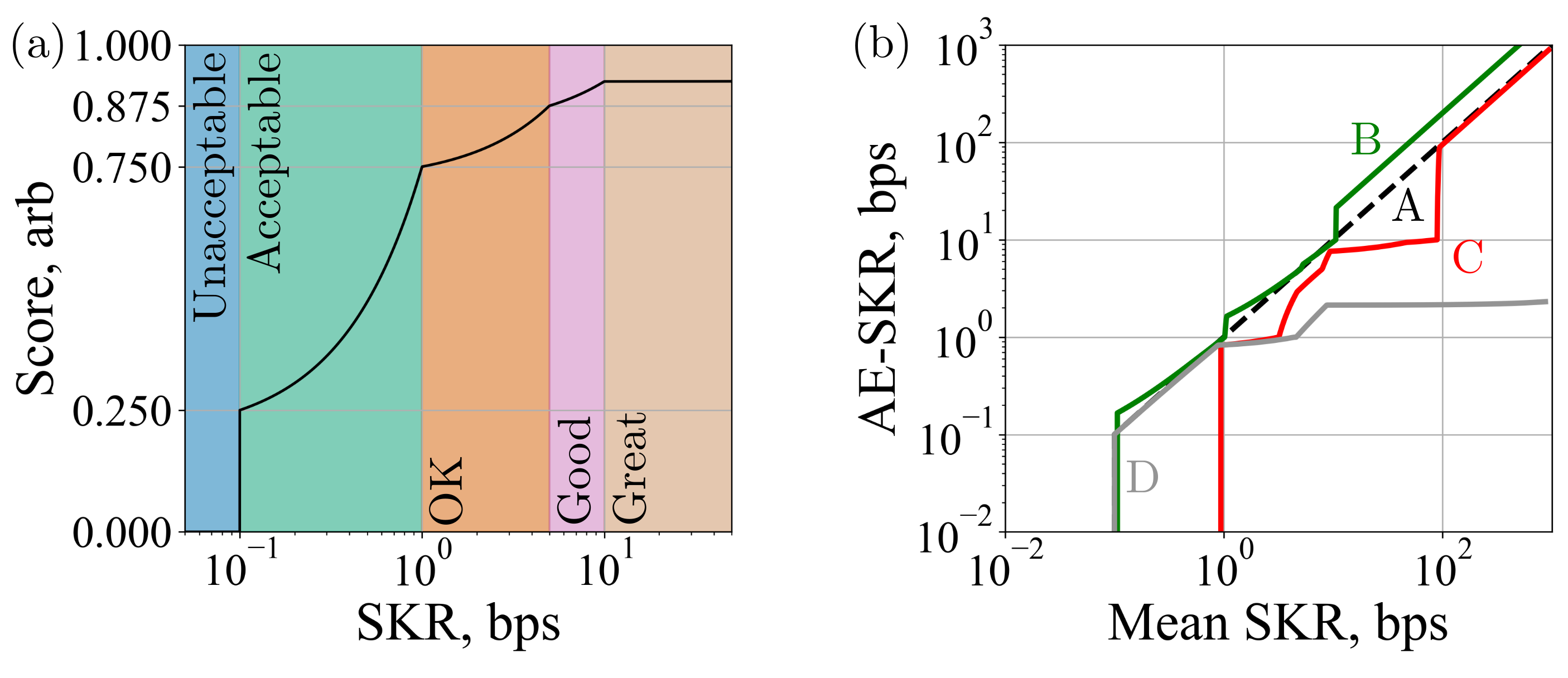}
    \caption{
    \textbf{(a)} shows the metric through which links in a network are scored according to the \acrfull{skr}. 
    There are 5 main sections. 
    The `Unacceptable' region, in blue, is where \gls{skr} $< 0.1$\,bps, where the \gls{skr} is deemed too low so the link is given the fail condition of $f($\gls{skr}$) = 0$. 
    The region of `Acceptable' \gls{skr} is where $0.1$\,bps $\le$ \gls{skr} $\le$ $1$\,bps, and is shown in green. The score in this region is linear where $0.25 \le f \le 0.75$. 
    A link that is labelled as `OK', in orange, has $1$\,bps $\le$ \gls{skr} $\le$ $5$\,bps and has linear score of $0.75 \le f($\gls{skr}$) \le 0.875$.
    The `Good' region, in pink, is where $5$\,bps $\le$ \gls{skr} $\le$ $10$\,bps, and has linear score of $0.875 \le f($\gls{skr}$) \le 0.925$.
    A `Great' link, shown in brown, has $10$\,bps $\le$ \gls{skr} $\le$ $10^{12}$\,bps, and  has linear score of $0.925 \le f($\gls{skr}$) \le 1$.
    A \gls{skr} of above $10^{12}$\,bps is set to $f($\gls{skr}$) = 1$, as no increase in \gls{skr} will have a real impact on communication speed in a \gls{qkd} network.
    \textbf{(b)} shows the relation between the mean of a set of 16 links to the \acrfull{aeskr}.
    Each show 16 communication links, where \textbf{A} has equal \gls{skr} in each link, \textbf{B} shows where 14 have equal \gls{skr} and the remaining 2 have \gls{skr} ten time higher than the rest, \textbf{C} shows where 14 have equal \gls{skr} and the remaining 2 have \gls{skr} ten time less than the rest, and \textbf{D} shows where 14 links have the same \gls{skr} and the final 2 links have a fixed \gls{skr} of $0.1$\,bps.
    }
    \label{fig:Score}
\end{figure}

\begin{figure}[ht]
    \centering
    \vspace{10pt}
    \includegraphics[width = 0.6\textwidth]{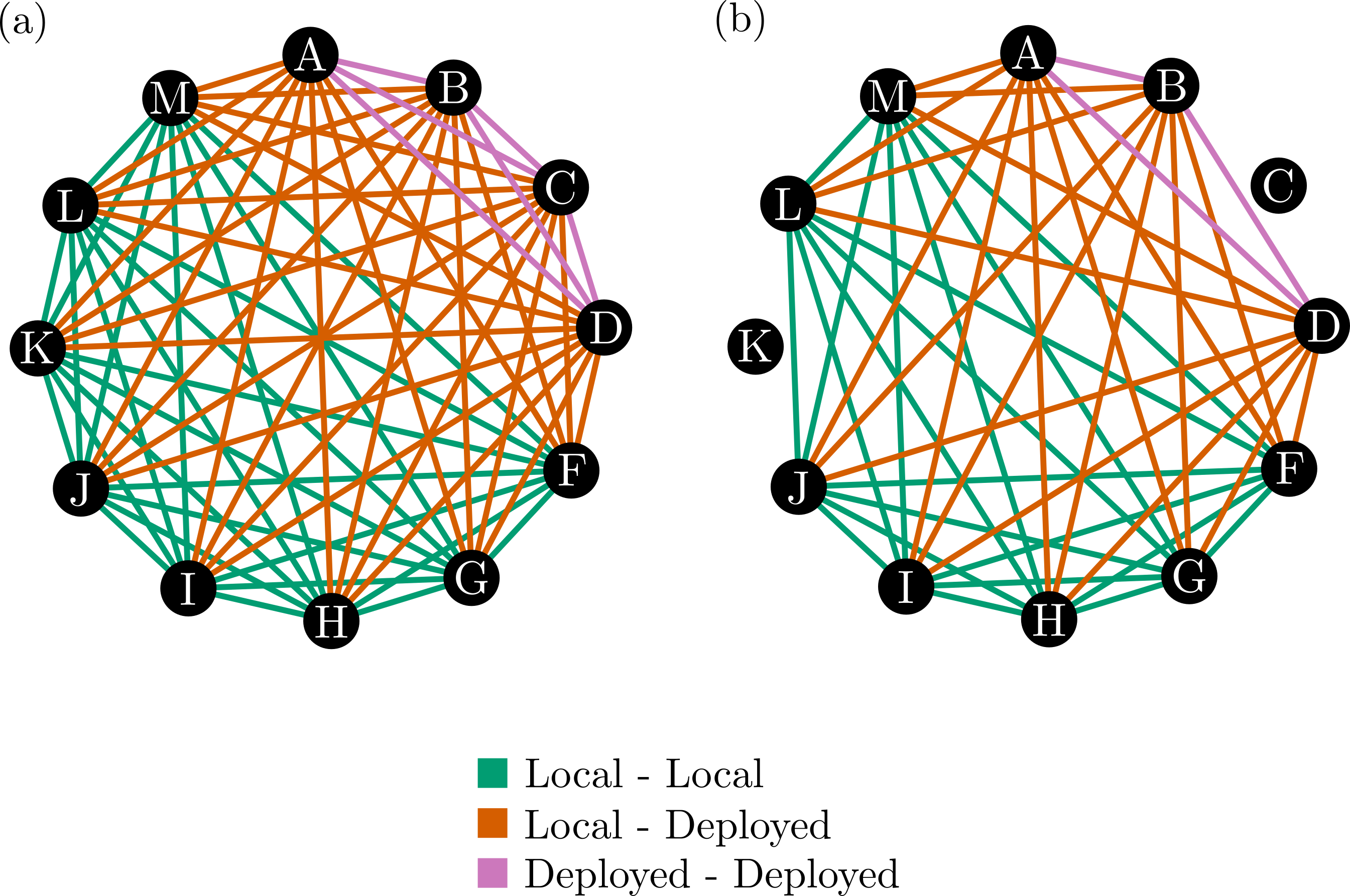}
    \caption{
    \textbf{(a)} shows the entanglement connectivity of a 12 user full mesh \gls{qn}.
    \textbf{(b)} shows the entanglement connectivity of the 10 user fully connected \gls{qn} discussed.
    These display the different combinations of local and deployed connections by colour.}
    \label{fig:Topology}
\end{figure}

Once separated, each split \gls{lc} travels though a separate \gls{fpc}, and finally each set of \glspl{lc} are combined into a single fibre.
A set of \glspl{lc} are selected for a single user to receive, such that they share entanglement with all other users in the network, as shown in Figure \ref{fig:Source}\,(e). 
This may be performed dynamically such that the connectivity of a network may be continuously changed.
Dynamicity allows for situations where an actively switched network would improve the performance aswell as \gls{lc} reassignment when network users are changed. 
The light from the single fibre is then measured by a user inside of a \gls{rq}. 
This measurement has a passive basis choice, preformed by a beam splitter, and then overlaps one of the two measurements from each basis onto one from the other basis reducing the number of \gls{spd} required at each user.
Once all users are receiving photons from the \gls{es}, correlations are calculated between every pair in the network, allowing users to generate a \gls{sk} through the Bennett-Brassard-Mermin \gls{qkd} protocol\cite{Bennett1992QuantumTheorem,Waks2002SecurityAttacks}, commonly known as the BBM92 protocol.



This infrastructure constructs a \gls{qn} consisting of 12 users with a full mesh topology, as shown in Figure \ref{fig:Topology}\,(a). 
There are four remote users; 
Alice, Bob, Chloe, and Dave
each connected through a deployed fibre link with bounce-back link losses of; $1.45$\,dB, $1,8$\,dB, $2.74$\,dB, and $3.24$\,dB respectively. 
The remaining eight users; 
Faye, Gopi, Heidi, Ivan, Jo, Kevin, Lea, and Marek
are connected through local fibre. 


To assess the quality of the network a metric is required.
The mean \glspl{skr} of a network often ignore links that are non-functional due to the increased \glspl{skr} of other links.
We select an interpolation function, as shown in figure \ref{fig:Score}\,(a).
This allows the choice of a failure condition in the network, here a link with \gls{skr} $< 0.1$\,bps, the score would be 0.
A network score is then given by;
\begin{equation}
    W = \sqrt[n]{ \prod_{i=1}^{n} f(r_i) },
    \label{eqn:W}
\end{equation}
where $n$ is the number of links in the \gls{qn}, $r_i$ is the \gls{skr} of link $i$, and $f$ is the interpolation function.
To retrieve an analogue to the \gls{qn} mean \gls{skr}, taking the score weighing into account, the inverse of $f$ can be applied to $W$.
This weighted mean is called the \gls{aeskr}.
The same process can be followed to generate the \gls{aeskr} of a subgroup of links, such as the connectivity scenario or for a single user.
Figure \ref{fig:Score}\,(b) shows how a small number of under-preforming link affects the scaling of a network \gls{aeskr}.

\section{RESULTS AND DISCUSSION}
\label{sec:results}

\begin{table}[b]
    \centering
    \vspace{0pt}
    \begin{tabular}{|l|c|}
         \hline
         \textbf{User} & \textbf{Average Visibility} \\ \hline \hline
         
         Alice & $0.99505$ \\ \hline
         Bob   & $0.99555$ \\ \hline
         Chloe & $0.99213$ \\ \hline
         Dave  & $0.98747$ \\ \hline
         Faye  & $0.99407$ \\ \hline
         Gopi  & $0.99552$ \\ \hline
         Heidi & $0.99533$ \\ \hline
         Ivan  & $0.99385$ \\ \hline
         Jo    & $0.99715$ \\ \hline
         Kevin & Failed \\ \hline 
         Lea   & $0.99347$ \\ \hline
         Marek & $0.99564$ \\ \hline

    \end{tabular}
    \vspace{5pt}
    \caption{\label{tab:userVisibility}
    The average polarisation visibility of \acrlong{lc}s being received by each user as measured from the \acrlong{pd}.
    }
\end{table}

\begin{figure}[t]
    \centering
    \includegraphics[width = 0.7\textwidth]{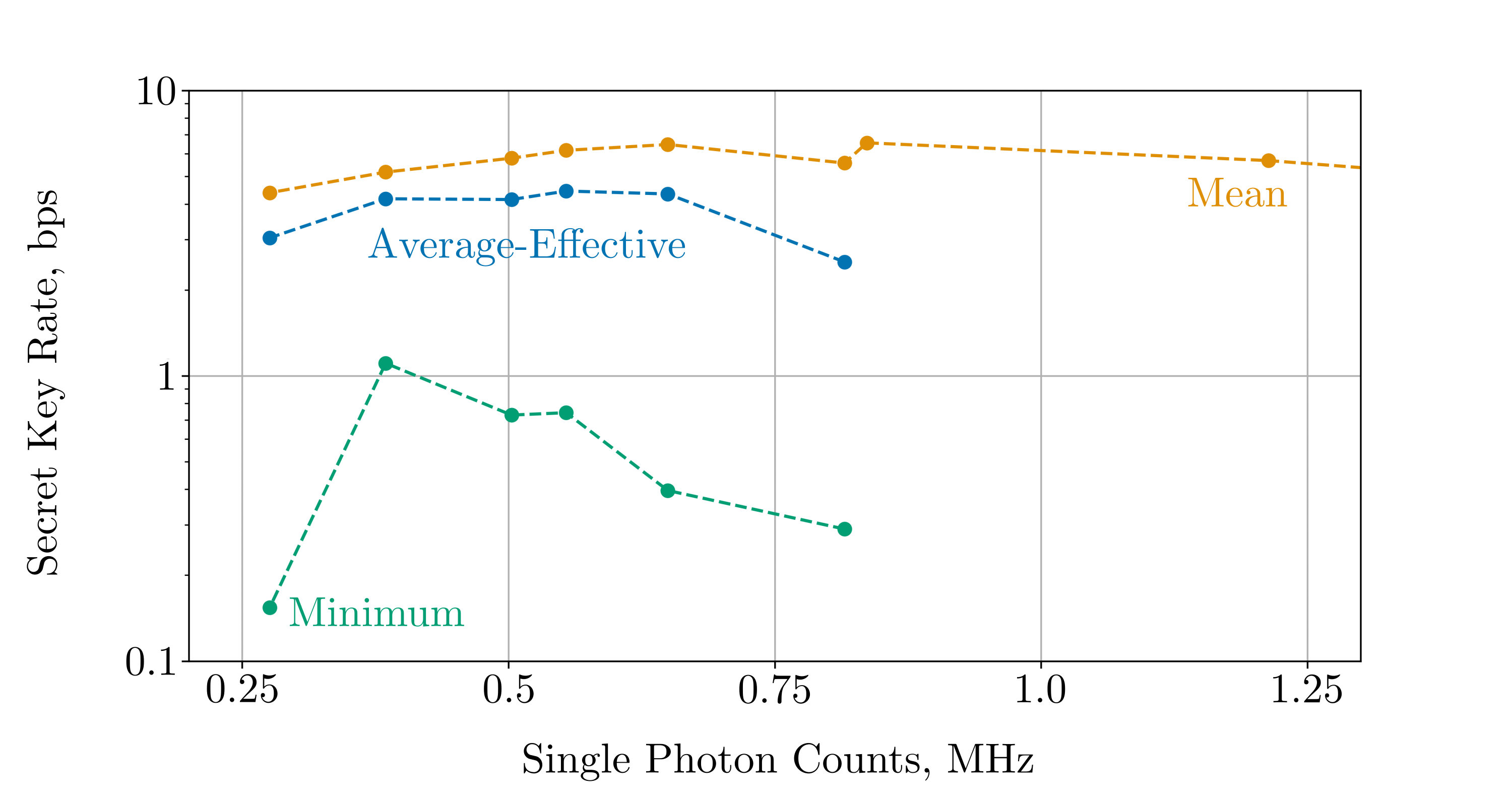}
    \caption{
    A sweep across the reference single photon rate of the system and the \acrfull{skr} of the system.
    The Mean \gls{skr} shows the average \gls{skr} across the \acrfull{qn} for the specified singles rate.
    The Minimum \gls{skr} shows the minimum \gls{skr} seen across the \gls{qn} for the specified singles rate.
    The Average-Effective \gls{skr} is calculated from the average score of each link, as shown in Figure \ref{fig:Score} (a) and given in Equation \ref{eqn:W}. 
    From this a singles rate of $\sim 0.45$\,MHz was selected for the \gls{qn}.
    }
    \label{fig:PumpSetting}
\end{figure}

In order to show entanglement, and therefore generate a \gls{sk}, between separated users with a shared \gls{es}, the definitions of polarisation must match between \glspl{rq}. 
To do this a canonical method of manual fibre neutralisation is used \cite{Peranic2022PolarizationNetworks}. 
This is performed by sending an attenuated classical signal comprised of two orthogonal polarisation definitions, \gls{h} and \gls{a} light, to the \glspl{rq} from the \acrlong{pd} setup, as shown in Figure \ref{fig:Source}\,(a). 
The average polarisation visibility of each user is provided in Table \ref{tab:userVisibility}.
It is clear that $11$ of the users have visibilities above $98.7 \%$, however Kevin failed the neutralisation steps with an estimated visibility of $\sim 97 \%$.
Due to the poor quality of the polarisation visibility, Kevin will not be included in the \gls{qn}.
However, it is key to note here that the estimated \gls{qber} of the Kevin Links were below 0.1, despite the measurement basis orthogonality issue with the \gls{rq}.
The timing jitter of the \glspl{spd} used by Chloe were significantly higher than other users.
As such Chloe is deemed to be a failed user and shall not be included in the \gls{qn}.
Although Chloe is a failed user, it is key to note that all links to other users could successfully generate a \gls{sk}.
This is despite the \gls{qber} being close to the maximum \gls{sk} generating \gls{qber}, but the \gls{skr} quickly dropped below the failure condition due to polarisation drifts.
The presented \gls{qn} is therefore a fully connected 10 user \glspl{qn}, as shown in Figure \ref{fig:Topology}\,(b), with no difference in resource allocation to a fully connected 12 user \gls{qn}, as shown in Figure \ref{fig:Topology}\,(a).

Once all users are receiving entangled pairs and the definitions of polarisation match, the optimal pump power must be found.
This is done by adjusting the laser power and measuring the number of single photon counts measured in an unused \gls{lc}.
Here the reference count rate is measured on \gls{lc} $+3$.
Figure \ref{fig:PumpSetting} shows the relation between the reference counts and the \gls{skr}.
Here it can be seen that the mean \gls{skr} remains high for much higher count rates, even past where the minimum \gls{skr} is zero.
The \gls{aeskr} is consistent between $0.38$\,MHz and $0.65$\,MHz, so a corrected count rate of $0.45$\,MHz was selected for the network.

\begin{figure}[p]
    \centering
    \includegraphics[width = 1\textwidth]{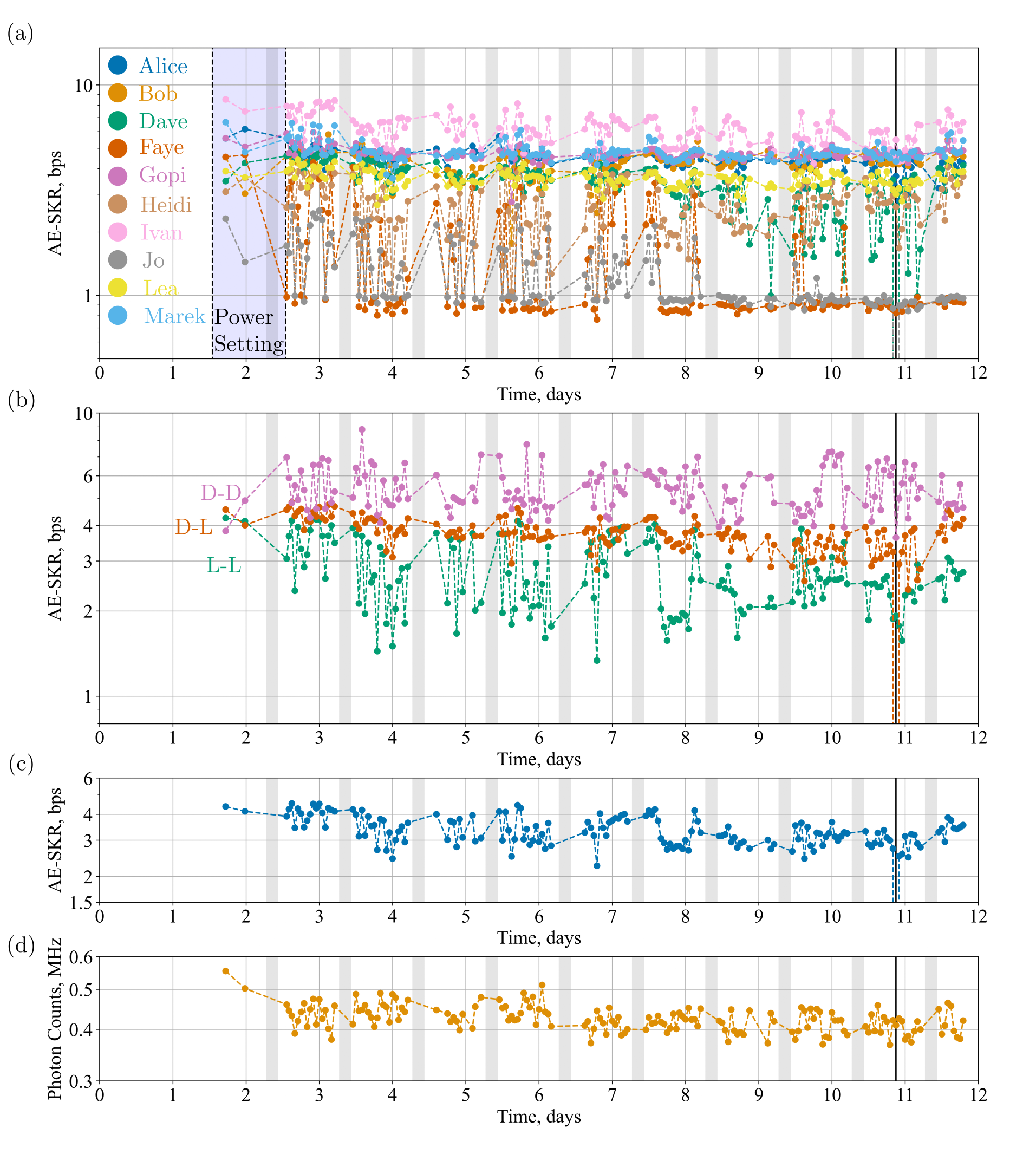}
    \caption{
    \textbf{(a)} shows the \acrfull{aeskr} for each user in the network.
    The blue box represents the time period in which the power per \acrfull{lc} was set.
    \textbf{(b)} shows the \gls{aeskr} each connectivity scenario.
    Here D-D indicates that both users are connected through deployed fibre, D-L indicates that one user is connected through deployed fibre and one is connected through local fibre, and L-L indicates that both users are connected through local fibre.
    \textbf{(c)} shows the \gls{aeskr} for the 10 user fully connected network, as shown in Figure \ref{fig:Topology}\,(b).
    \textbf{(d)} shows the count rate of the reference \gls{lc} corrected for loss.
    All data points in this figure is an average over 10 minutes, and grey boxes indicate when the \acrlong{spd}s stop functioning to cycle the cryogenic gasses.
    }
    \label{fig:Stability}
\end{figure}

When the Neutralisation of all \gls{qn} links is complete network up-time is defined to have started. 
The count rate of the \glspl{lc} are set between $1.5$ and $2.5$\,days, as shown by the blue region in Figure \ref{fig:Stability}\,(a).
Two data points, closest to the selected pump power, are shown in this region.
Most users \gls{aeskr} stays stable across the whole time range.
Some users; Faye, Heidi, and Jo, fluctuate in \gls{aeskr} after setup but stabilise after some time.
One user, Dave, fluctuates across the whole time set and then the link with user Jo fails after $10.8$\, days of network up-time.
This point is the Network Failure point, shown in Figure \ref{fig:Stability} as as the vertical black line, and the network would be reset.
Notably, even though the network was deemed to have failed, the failure was reached by one link dropping to $0.096$\,bps.
As can be seen in Figure \ref{fig:Stability}\,(c), the network recovers functionality after the failure condition was reached.

Figure \ref{fig:Stability}\,(b) shows the three connectivity scenarios presented in Figure \ref{fig:Topology}\,(b). 
The scenario with the highest \gls{aeskr} was when both users are connected though deployed fibre, with $5.04$\,bps, as shown in Table \ref{tab:userAESKR}.
This occurred as the detectors with the highest efficiency and \glspl{rq} with the lowest loss were given to the deployed users, therefore higher transmission loss is balanced with the more efficient measurement.
The scenario where the \gls{aeskr} was lowest was when both users were connected with local fibre.
The stability of all three scenarios is similar with approximately $50\%$ fluctuation from the full time \gls{aeskr} to the minimum and maximum \gls{aeskr}.
This demonstrates that across the whole $10.8$\,days there was a stable and fully functional $10$ user fully connected \gls{qn}.

\begin{table}[ht]
    \centering
    \vspace{5pt}
    \begin{tabular}{|l|c|c|c|}
         \hline
         
         \textbf{Scenario} & \textbf{AE-SKR} & \textbf{Maximum} & \textbf{Minimum} \\ \hline \hline
         
         Alice              & $4.4983$ & $6.1654$ & $3.2324$ \\ \hline
         Bob                & $4.2148$ & $5.8158$ & $1.7643$ \\ \hline
         Dave               & $3.4391$ & $4.9570$ & $0.9971$ \\ \hline
         Faye               & $0.9578$ & $4.6524$ & $0.7678$ \\ \hline
         Gopi               & $4.6861$ & $5.9627$ & $2.7813$ \\ \hline
         Heidi              & $2.7457$ & $4.0443$ & $0.9795$ \\ \hline
         Ivan               & $6.1746$ & $8.5474$ & $4.2368$ \\ \hline
         Jo                 & $0.9874$ & $2.6596$ & $0.8514$ \\ \hline
         Lea                & $3.6083$ & $4.4097$ & $2.6903$ \\ \hline
         Marek              & $4.7446$ & $6.6571$ & $3.6015$ \\ \hline \hline
         
         Local-Local        & $2.7706$ & $4.3851$ & $1.3378$ \\ \hline
         Local-Deployed     & $3.7769$ & $4.8376$ & $2.5522$ \\ \hline
         Deployed-Deployed  & $5.0436$ & $8.7522$ & $3.8343$ \\ \hline \hline
         
         Full Network       & $3.3818$ & $4.5224$ & $2.2564$ \\ \hline

    \end{tabular}
    \vspace{5pt}
    \caption{\label{tab:userAESKR}
    The \acrfull{aeskr} of each user, in bps, of each scenario given in Figure \ref{fig:Stability}, before the network failed at $10.8$ days.
    The maximum and minimum are the maximum \gls{aeskr} and minimum \gls{aeskr} respectively, before the network failed.
    }
\end{table}

\section{CONCLUSION}
\label{sec:discussion}
Here we have demonstrated the setup and operation of a polarisation based 10 user fully connected \gls{ed} \gls{qn}, with the resources for a 12 user fully connected \gls{ed} \gls{qn}.
The \gls{qn} demonstrated a minimum polarisation stability duration of $10.8$\,days through both local and deployed fibre links with network wide \gls{aeskr} of $3.3818$\,bps, as shown in Table \ref{tab:userAESKR}.
Previously, \gls{ed} \glspl{qn} based on entangled photon pairs with passive polarisation compensation, with deployed fibre links have only shown up to a day of continuous polarisation stability before the \gls{qn} is reset \cite{Joshi2020ANetwork}.
It is also shown that the links between users with deployed fibres demonstrate similar stability to users with only local fibre links.
This improvement in stability indicates that the limitation of metropolitan scale \gls{ed} \glspl{qn} is likely to be the polarisation stability of the distribution devices 
rather than the deployed fibre links.
The large concentration of fibre lengths that sit in a relatively small area within the distribution devices leads to large polarisation drifts from small environment drifts.
Metropolitan scale deployed fibres do not have a large influence on the stability of a \gls{qn}, and as such scaling \glspl{qn} up to arbitrary deployed links should be possible, to the limit of the fibre loss of each link.


\acknowledgments 
 

This work was supported by the UK Engineering and Physical Sciences Research Council (EPSRC) grants EP/SO23607/1, for the Quantum Engineering Centre for Doctoral Training, Centre for Nanoscience \& Quantum Information, University of Bristol, and EP/T001011/1 for the Quantum Communications Hub. This work was supported by the Ministry of Science and Education (MSE) of Croatia [contract No. KK.01.1.1.01.0001]

\bibliography{report} 

\begin{thebibliography}{10}

\bibitem{Joshi2020ANetwork}
Joshi, S.~K., Aktas, D., Wengerowsky, S., Loncaric, M., Neumann, S.~P., Liu,
  B., Scheidl, T., Lorenzo, G.~C., Samec, Z., Kling, L., Qiu, A., Razavi, M.,
  Stipcevic, M., Rarity, J.~G., and Ursin, R., ``{A trusted node-free
  eight-user metropolitan quantum communication network},'' {\em Science
  Advances}~{\bf 6},  eaba0959 (9 2020).

\bibitem{Wang2022ANetwork}
Wang, R., Alia, O., Clark, M.~J., Bahrani, S., Joshi, S.~K., Aktas, D.,
  Kanellos, G.~T., Perani{\'{c}}, M., Lon{\v{c}}ari{\'{c}}, M.,
  Stip{\v{c}}evi{\'{c}}, M., Rarity, J., Nejabati, R., and Simeonidou, D., ``{A
  Dynamic Multi-Protocol Entanglement Distribution Quantum Network},'' in [{\em
  Optical Fiber Communication Conference (OFC)}{\nolinebreak\hspace{0.1em}]},
  paper Th3D.3, Optica Publishing Group (3 2022).

\bibitem{Huang2022ExperimentalNetwork}
Huang, Z., Joshi, S.~K., Aktas, D., Lupo, C., Quintavalle, A.~O.,
  Venkatachalam, N., Wengerowsky, S., Lon{\v{c}}ari{\'{c}}, M., Neumann, S.~P.,
  Liu, B., Samec, Z., Kling, L., Stip{\v{c}}evi{\'{c}}, M., Ursin, R., and
  Rarity, J.~G., ``{Experimental implementation of secure anonymous protocols
  on an eight-user quantum key distribution network},'' {\em npj Quantum
  Information 2022 8:1}~{\bf 8},  1--7 (3 2022).

\bibitem{Pelet2022UnconditionallyNetwork}
Pelet, Y., Puthoor, I.~V., Venkatachalam, N., Wengerovsky, S., Loncaric, M.,
  Neumann, S.~P., Liu, B., Samec, Z., Stip{\v{c}}evi{\'{c}}, M., Ursin, R.,
  Andersson, E., Rarity, J.~G., Aktas, D.~V., and Joshi, S.~K.,
  ``{Unconditionally secure digital signatures implemented in an eight-user
  quantum network*},'' {\em New Journal of Physics}~{\bf 24},  093038 (10
  2022).

\bibitem{Wehner2018QuantumAhead}
Wehner, S., Elkouss, D., and Hanson, R., ``{Quantum internet: A vision for the
  road ahead},'' (10 2018).

\bibitem{Chung2022DesignNetwork}
Chung, J., Eastman, E.~M., Kanter, G.~S., Kapoor, K., Lauk, N., Pe{\~{n}}a, C.,
  Plunkett, R., Sinclair, N., Thomas, J.~M., Valivarthi, R., Xie, S.,
  Kettimuthu, R., Kumar, P., Spentzouris, P., and Spiropulu, M., ``{Design and
  Implementation of the Illinois Express Quantum Metropolitan Area Network},''
  {\em arxiv:2207.09589 [quant-ph]}  (7 2022).

\bibitem{Chen2010MetropolitanNetwork}
Chen, T.-Y., Wang, J., Liang, H., Liu, W.-Y., Liu, Y., Jiang, X., Wang, Y.,
  Wan, X., Cai, W.-Q., Ju, L., Chen, L.-K., Wang, L.-J., Gao, Y., Chen, K.,
  Peng, C.-Z., Chen, Z.-B., and Pan, J.-W., ``{Metropolitan all-pass and
  inter-city quantum communication network},'' {\em Optics Express}~{\bf 18},
  27217 (12 2010).

\bibitem{Tang2022Free-runningDistribution}
Tang, B.-Y., Chen, H., Wang, J.-P., Yu, H.-C., Shi, L., Sun, S.-H., Peng, W.,
  Liu, B., and Yu, W.-R., ``{Free-running long-distance
  reference-frame-independent quantum key distribution},'' {\em npj Quantum
  Information 2022 8:1}~{\bf 8},  1--8 (9 2022).

\bibitem{Chen2021AnKilometres}
Chen, Y.~A., Zhang, Q., Chen, T.~Y., Cai, W.~Q., Liao, S.~K., Zhang, J., Chen,
  K., Yin, J., Ren, J.~G., Chen, Z., Han, S.~L., Yu, Q., Liang, K., Zhou, F.,
  Yuan, X., Zhao, M.~S., Wang, T.~Y., Jiang, X., Zhang, L., Liu, W.~Y., Li, Y.,
  Shen, Q., Cao, Y., Lu, C.~Y., Shu, R., Wang, J.~Y., Li, L., Liu, N.~L., Xu,
  F., Wang, X.~B., Peng, C.~Z., and Pan, J.~W., ``{An integrated
  space-to-ground quantum communication network over 4,600 kilometres},'' {\em
  Nature 2020 589:7841}~{\bf 589},  214--219 (1 2021).

\bibitem{Neumann2022ContinuousLink}
Neumann, S.~P., Buchner, A., Bulla, L., Bohmann, M., and Ursin, R.,
  ``{Continuous entanglement distribution over a transnational 248 km fiber
  link},'' {\em Nature Communications 2022 13:1}~{\bf 13},  1--8 (10 2022).

\bibitem{Pompili2021RealizationQubits}
Pompili, M., Hermans, S.~L., Baier, S., Beukers, H.~K., Humphreys, P.~C.,
  Schouten, R.~N., Vermeulen, R.~F., Tiggelman, M.~J., dos Santos~Martins, L.,
  Dirkse, B., Wehner, S., and Hanson, R., ``{Realization of a multinode quantum
  network of remote solid-state qubits},'' {\em Science}~{\bf 372},  259--264
  (4 2021).

\bibitem{Avis2022AnalysisNode}
Avis, G., Rozp{\c{e}}dek, F., and Wehner, S., ``{Analysis of Multipartite
  Entanglement Distribution using a Central Quantum-Network Node},'' (3 2022).

\bibitem{Bennett1992QuantumTheorem}
Bennett, C.~H., Brassard, G., and Mermin, N.~D., ``{Quantum cryptography
  without Bell’s theorem},'' {\em Physical Review Letters}~{\bf 68},  557 (2
  1992).

\bibitem{Waks2002SecurityAttacks}
Waks, E., Zeevi, A., and Yamamoto, Y., ``{Security of quantum key distribution
  with entangled photons against individual attacks},'' {\em Physical Review
  A}~{\bf 65},  052310 (4 2002).

\bibitem{Peranic2022PolarizationNetworks}
Perani{\'{c}}, M., Clark, M., Wang, R., Bahrani, S., Alia, O., Wengerowsky, S.,
  Radman, A., Lon{\v{c}}ari{\'{c}}, M., Stip{\v{c}}evi{\'{c}}, M., Rarity, J.,
  Nejabati, R., and Joshi, S.~K., ``{Polarization compensation methods for
  quantum communication networks},'' {\em arxiv:2208.13584 [quant-ph]}  (8
  2022).

\end{thebibliography}
\bibliographystyle{spiebib} 



\end{document}